\RequirePackage{lineno} 
\documentclass[aps,prc,superscriptaddress,showpacs,floatfix,nofootinbib,twocolumn]{revtex4}
\usepackage{graphicx} 
\usepackage{float} 
\usepackage{amssymb}
\usepackage{url}
\usepackage{harpoon}
\usepackage{amsmath}
\usepackage{MnSymbol}
\usepackage{multirow}
\usepackage[colorlinks,breaklinks]{hyperref}
\hypersetup{linkcolor=blue,citecolor=blue,filecolor=black,urlcolor=blue}
\newcommand{\pT} {\ensuremath{p_{\mathrm{T}}}}

\newcommand{\vns}{v_{n}^{\mathrm{s}}}
\newcommand{\vnc}{v_{n}^{\mathrm{c}}}
\newcommand{\vnsr}{v_{n}^{\mathrm{s,raw}}}
\newcommand{\vncr}{v_{n}^{\mathrm{c,raw}}}

\newcommand{\npartf}{N_{\mathrm {part}}^{\mathrm{F}}}
\newcommand{\npartb}{N_{\mathrm {part}}^{\mathrm{B}}}
\newcommand{\npart}{N_{\mathrm {part}}}

\newcommand{\vtwos}{v_{2}^{\mathrm{s}}}
\newcommand{\vtwoc}{v_{2}^{\mathrm{c}}}
\newcommand{\vthrs}{v_{3}^{\mathrm{s}}}
\newcommand{\vthrc}{v_{3}^{\mathrm{c}}}

\begin{document}
\title{Forward-backward eccentricity and participant-plane angle fluctuations and their influences on longitudinal dynamics of collective flow }
\newcommand{\sunysb}{Department of Chemistry, Stony Brook University, Stony Brook, NY 11794, USA}
\newcommand{\bnl}{Physics Department, Brookhaven National Laboratory, Upton, NY 11796, USA}
\author{Jiangyong Jia}\email[Correspond to\ ]{jjia@bnl.gov}
\affiliation{\sunysb}\affiliation{\bnl}
\author{Peng Huo}\affiliation{\sunysb}

\begin{abstract}
We argue that the transverse shape of the fireball created in heavy ion collision is controlled by event-by-event fluctuations of the eccentricity vectors for the forward-going and backward-going wounded nucleons: $\vec{\epsilon}_n^{\mathrm{F}}\equiv \epsilon_n^{\mathrm{F}} e^{i n\Phi_n^{\mathrm{*F}}}$ and $\vec{\epsilon}_n^{\mathrm{B}}\equiv \epsilon_n^{\mathrm{B}} e^{i n\Phi_n^{\mathrm{*B}}}$. Due to the asymmetric energy deposition of each wounded nucleon along its direction of motion, the eccentricity vector of the produced fireball is expected to interpolate between $\vec{\epsilon}_n^{\mathrm{F}}$ and $\vec{\epsilon}_n^{\mathrm{B}}$ along the pseudorapidity, and hence exhibits sizable forward-backward(FB) asymmetry ($\epsilon_n^{\rm B}\neq\epsilon_n^{\rm F}$) and/or FB-twist ($\Phi_n^{\mathrm{*F}}\neq\Phi_n^{\mathrm{*B}}$). A transport model calculation shows that these initial state longitudinal fluctuations for $n=2$ and 3 survive the collective expansion, and result in similar FB asymmetry and/or a twist in the final state event-plane angles. These novel EbyE longitudinal flow fluctuations should be accessible at RHIC and the LHC using the event-shape selection technique proposed in earlier papers. If these effects are observed experimentally, it could improve our understanding of the initial state fluctuations, particle production and collective expansion dynamics of the heavy ion collision.
\end{abstract}
 \maketitle 

\section{Introduction} 
\label{sec:1}
Relativistic heavy-ion collisions at RHIC and the LHC produce a quark-gluon fireball, whose shape is lumpy and asymmetric in the transverse plane~\cite{Gyulassy:1996br}. The fireball expands under large pressure gradients, which transfer the inhomogenous initial condition into azimuthal anisotropy of produced particles in momentum space~\cite{Ollitrault:1992bk,Alver:2010gr}. Hydrodynamic models are used to understand the space-time evolution of the fireball from the measured azimuthal anisotropy~\cite{Gale:2013da,Heinz:2013th,Luzum:2013yya}. The success of these models in describing the anisotropy of particle production in heavy-ion collisions at RHIC and the LHC~\cite{Adare:2011tg,Adamczyk:2013waa,ALICE:2011ab,Aad:2012bu,Chatrchyan:2013kba,Aad:2013xma,Aad:2014fla} places important constraints on the transport properties and initial conditions of the produced matter~\cite{Luzum:2012wu,Teaney:2010vd,Gale:2012rq,Niemi:2012aj,Qiu:2012uy,Teaney:2013dta}.

When describing the dynamics of the fireball in azimuthal angle $\phi$, it is convenient to parameterize both the initial shape and final state anisotropy of the fireball in terms of a Fourier decomposition. Various shape components of the fireball are described by the eccentricities $\epsilon_n$ and participant-plane (PP) angle $\Phi_n^*$, calculated from transverse positions $(r,\phi)$ of the participating nucleons relative to their center of mass~\cite{Alver:2010gr,Teaney:2010vd}:
\begin{eqnarray}
\label{eq:1}
\vec{\epsilon}_n\equiv \epsilon_n e^{i n\Phi_n^*} \equiv -\frac{\langle r^n e^{i n\phi} \rangle}{\langle r^n\rangle}\;.
\end{eqnarray} 
The azimuthal anisotropy in the distribution of produced particles is expressed as:
\begin{equation}
\label{eq:2}
dN/d\phi\propto1+2\sum_{n=1}^{\infty}v_{n}\cos n(\phi-\Phi_{n})\;,
\end{equation}
where $v_n$ and $\Phi_n$ represent the magnitude and phase (referred to as the event plane or EP) of the $n^{\mathrm{th}}$-order harmonic flow. They are also written in a compact form:
\begin{eqnarray}
\label{eq:2b} 
\vec{v}_n\equiv v_n e^{i n\Phi_n}\;.
\end{eqnarray} 
According to hydrodynamic model calculations, elliptic flow $\vec{v}_2$ and triangular flow $\vec{v}_3$ are the dominant harmonics, and they are driven mainly by the ellipticity vector $\vec{\epsilon}_2$ and triangularity vector $\vec{\epsilon}_3$ of the initially produced fireball~\cite{Qiu:2011iv,Gardim:2011xv}. The origin of higher-order ($n>3$) harmonics is more complicated; they arise from both $\vec{\epsilon}_n$ and final state non-linear mixing of lower-order harmonics~\cite{Gardim:2011xv,Teaney:2012ke,Teaney:2013dta}. 

The connection between initial shape of the fireball and its final state harmonic flow is also applicable in the longitudinal direction. Although the rapidity distributions of the particle multiplicity exhibits forward-backward (FB) symmetry and boost-invariance near mid-rapidity when averaging over many events, this is not necessarily the case on an event-by-event (EbyE) basis~\cite{Bialas:2011bz,Bzdak:2012tp}: Particles in the forward (backward) rapidity are preferably produced by the participants in the forward-going (backward-going) nucleus, and the number of forward-going and backward-going participating nucleons are not the same in a given event, $\npartf\neq\npartb$~\cite{Bialas:1976ed}~\footnote{Responsible also for the FB-asymmetry of the multiplicity distribution in p+A collisions.}. Indeed, extensive analyses of experimental data~\cite{Back:2006id,Abelev:2009ag} reveal a strong FB-asymmetry in the particle production in pseudorapidity $\eta$~\cite{Bialas:2007eg,Bzdak:2009dr,Bialas:2011bz}.

In this work, we generalize the EbyE FB-asymmetry idea to the study of the rapidity fluctuations and event-plane decorrelation of harmonic flow. We first note that eccentricity vectors can be calculated separately for the two colliding nuclei from their respective participants via Eq.~\ref{eq:1}: $\vec{\epsilon}_n^{\mathrm{F}}\equiv \epsilon_n^{\mathrm{F}} e^{i n\Phi_n^{\mathrm{*F}}}$ and $\vec{\epsilon}_n^{\mathrm{B}}\equiv \epsilon_n^{\mathrm{B}} e^{i n\Phi_n^{\mathrm{*B}}}$. This together with the asymmetric energy deposition of participating nucleons suggests that the transverse shape of the initially produced fireball at the time of the thermalization but before the onset of the hydrodynamics should be a strong function of $\eta$. Consequently, the eccentricity vector that drives the evolution of the whole system, $\vec{\epsilon}_n^{\mathrm{tot}}$, should also depend on $\eta$. The value of $\vec{\epsilon}_n^{\mathrm{tot}}(\eta)$ is expected to interpolate between $\vec{\epsilon}_n^{\mathrm{F}}$ at forward rapidity and $\vec{\epsilon}_n^{\mathrm{B}}$ at the backward rapidity, and is equal to $\vec{\epsilon}_n$ only at mid-rapidity. Just as $\npartf\neq\npartb$ being the origin of FB multiplicity fluctuations, FB fluctuations of the eccentricities and PP angles, i.e. $\epsilon_n^{\mathrm{F}}\neq\epsilon_n^{\mathrm{B}}$ and $\Phi_n^{\mathrm{*F}}\neq\Phi_n^{\mathrm{*B}}$ in an event, are expected to give rise to significant the fluctuations of $\vec{\epsilon}_n^{\mathrm{tot}}(\eta)$ as a function of pesudorapidity. If we define
\begin{eqnarray}
\label{eq:3}
A_{\epsilon_n} &=& \frac{\epsilon_n^{\mathrm B}-\epsilon_n^{\mathrm F}}{\epsilon_n^{\mathrm B}+\epsilon_n^{\mathrm F}}\\\label{eq:3b}
A_{\mathrm{N_{part}}} &=& \frac{\npartb-\npartf}{\npartb+\npartf}
\end{eqnarray} 
to characterize the FB eccentricity and multiplicity asymmetries, then it is easy to show that $A_{\epsilon_n}\gg A_{N_{\mathrm{part}}}$ in Pb+Pb or Au+Au collisions. Furthermore, the twist angle, $\Phi_n^{\mathrm{*F}}-\Phi_n^{\mathrm{*B}}$, is also quite large, especially in very central and peripheral collisions where the shape fluctuations are large. Similar twist effects have been studied previously~\cite{Bozek:2010vz,Jia:2014vja} and are found to survive the hydrodynamic evolution~\cite{Bozek:2010vz}.

The FB eccentricity vector fluctuations are generic initial state long-range effects, and should be present as long as the particle production for each participating nucleon exhibit FB asymmetry. Since the hydrodynamic response to eccentricity vectors are known to be linear for elliptic flow and triangular flow, these FB eccentricity fluctuations should result in $\eta$-dependent flow with strong EbyE variations: $\vec{v}_2(\eta)\propto\vec{\epsilon}^{\mathrm{tot}}_2(\eta)$ and $\vec{v}_3(\eta)\propto\vec{\epsilon}^{\mathrm{tot}}_3(\eta)$. They should also lead to a new class of phenomena not yet explored experimentally, such as FB-asymmetry of harmonic flow, event-shape twist and non-trivial $\eta$-dependent mode-mixing effects~\cite{Schenke:2012fw,Huo:2013qma,Jia:2014vja}. Recently, measurements of two-particle correlations in high-multiplicity p+Pb and d+Au collisions suggest that hydrodynamic behavior may also be present in these small collision systems~\cite{CMS:2012qk,Abelev:2012ola,Aad:2012gla}. It would be interesting to study the longitudinal collective dynamics in these collisions, where FB eccentricity vector fluctuations are expected to be much bigger than for A+A collisions. 

In this paper, we present a study of the FB eccentricity vector fluctuations and their influence on the rapidity fluctuations and event-plane decorrelations of harmonic flow. Our intention is not to perform an exhaustive study over the full parameter space in $\npart,\npartf,\npartb,\epsilon_n,\epsilon_n^{\mathrm{F}},\epsilon_n^{\mathrm{B}},\Phi_n^*,\Phi_n^{\mathrm{*F}},\Phi_n^{\mathrm{*B}}$. Instead, we focus our attention on several classes of events with specific characteristics in these parameters and study $v_n(\eta)$ and $\Phi_n(\eta)$ in the final state. Such event-shape selection technique~\cite{Schukraft:2012ah,Huo:2013qma,Jia:2014vja} has the advantage of exposing certain aspect of the initial geometry fluctuations, and the signal remains large and easy to interpret after an ensemble average. We perform these studies using the AMPT transport model~\cite{Lin:2004en}, which contains both FB eccentricity fluctuations and collective flow. 

The structure of the paper is as follows. The next section presents a simple model, which establishes the magnitudes and general trends of flow fluctuations along $\eta$. Section~\ref{sec:3} discusses the classification of event shape in the AMPT model and introduces the relevant experimental observables. Sections~\ref{sec:4} and \ref{sec:5} present the results of $v_n(\eta)$ and event plane $\Phi_n(\eta)$ for several event classes. This paper ends with a discussion of the implication of these results and a conclusion.

\section{A simple model} 
\label{sec:2}
A simple model is used to describe the FB asymmetry in the shape of the initial fireball and how this asymmetry affects the flow harmonics. The model is based on a wounded-nucleon model~\cite{Bialas:1976ed} and an independent emission picture~\cite{Bozek:2010vz}, where the energy density profile and particle production along pseudorapidity in one event are related to the density profile of participating nucleons in each nuclei:
\begin{eqnarray}
\label{eq:4}
\rho(x,y,\eta) &=&  f^{\rm F}(\eta)\rho_{\rm F}(x,y)+ f^{\rm B}(\eta)\rho_{\rm B}(x,y)\\
dN/d\eta &\propto& f^{\rm F}(\eta)\npartf+f^{\rm B}(\eta)\npartb\;. 
\end{eqnarray} 
Here the $f^{\rm F}(\eta)$ or $f^{\rm B}(\eta)$ is the normalized emission function of one forward-going or backward-going wounded nucleon ($\int f^{\rm F}(\eta) d\eta=1$ and $\int f^{\rm B}(\eta) d\eta=1$), and $f^{\rm F}(\eta)=f^{\rm B}(-\eta)$ for symmetric collision system. One example emission function derived from RHIC data can be found in Ref.~\cite{Bialas:2007eg}. The distributions $\rho_{\rm B}$ and $\rho_{\rm F}$ are density profiles of participating nucleons in the two nuclei: $\int \rho_{\rm F}(x,y) dxdy=\npartf$ and $\int \rho_{\rm B}(x,y) dxdy =\npartb$. 

The shape of transverse density profile just after the collisions, when most entropy has been produced but before onset of hydrodynamics, should be $\eta$ dependent. According to Eq.~\ref{eq:4}, this $\eta$-dependent profile in a given event is naturally related to the eccentricity vectors of the two colliding nuclei:
\begin{eqnarray}
\label{eq:5}
\vec{\epsilon}^{\mathrm{tot}}_n(\eta)\approx \alpha(\eta)\vec{\epsilon}_n^{\rm F}+(1-\alpha(\eta))\vec{\epsilon}_n^{\rm B}\equiv\epsilon^{\mathrm{tot}}_n(\eta)e^{in\Phi_n^{*\rm tot}(\eta)}.
\end{eqnarray} 
where~\footnote{The center-of-mass of the participants in the two nuclei in general can be different, leading to a correction to Eq.~\ref{eq:5} around mid-rapidity. This correction can be significant for $\vec{\epsilon}_2$ (Fig.~\ref{fig:2} (a) and Appendix~\ref{sec:app1}) or when $\npartf$ or $\npartb$ are small, such as in peripheral collisions or asymmetric collisions. This effect is ignored in this discussion.} 
\begin{eqnarray}
\label{eq:5b}
\alpha(\eta) =\frac{ f^{\rm F}(\eta) \npartf \langle r^n\rangle^{\rm F}}{f^{\rm F}(\eta) \npartf \langle r^n\rangle^{\rm F}+f^{\rm B}(\eta) \npartb \langle r^n\rangle^{\rm B}}\;,\end{eqnarray} 
is the $\eta$ dependent weighing factor for forward-going participating nucleons. The value of $\alpha$ is determined by the emission profiles, but also depends on the number and the transverse profile of participating nucleons in each nuclei via $\npart$ and $\langle r^n\rangle$. It is easy to see that $\alpha(-\infty)=0$ and $\alpha(\infty)=1$, and it's value fluctuates EbyE around 1/2 at mid-rapidity for a symmetric collision system, hence $\vec{\epsilon}^{\mathrm{tot}}_n(0)\approx\vec{\epsilon}_n$.

\begin{figure}[!t]
\begin{center}
\includegraphics[width=1\columnwidth]{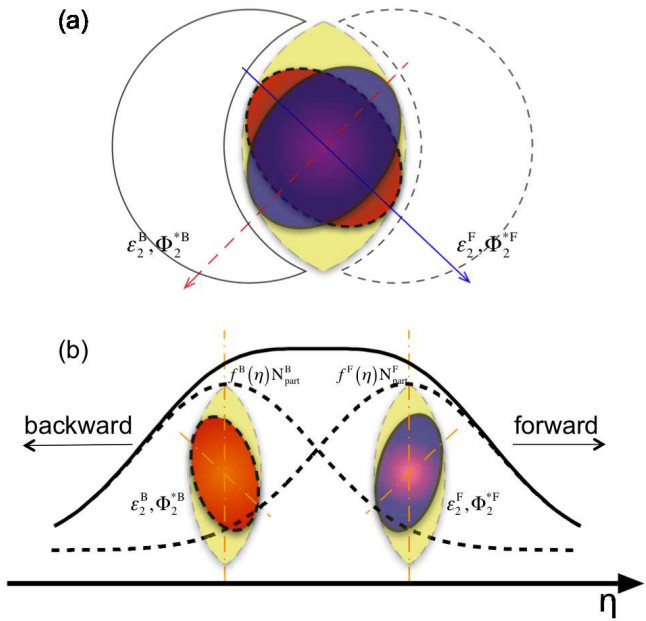}
\end{center}
\caption{\label{fig:idea} Schematic illustration of the forward-backward fluctuation of second-order eccentricity and participant plane, in transverse plane (a) and along rapidity direction (b) in A+A collisions. The dashed-lines indicate the particle production profiles for forward-going and backward-going participants, $f^{\rm F}(\eta)\npartf$ and $f^{\rm B}(\eta)\npartb$, respectively.}
\end{figure}

Figure~\ref{fig:idea} illustrates the origin of the $\eta$-dependence of the eccentricity implied by Eq.~\ref{eq:5}, which is the main idea of this paper.
Several conclusions can be drawn from this equation. First, if harmonic flow at a given $\eta$ is driven by the corresponding eccentricity vector at the same $\eta$, which is a reasonable assumption for $n=2$ and 3~\cite{Qiu:2011iv,Gardim:2011xv},  we should expect the following relation to hold:
\begin{eqnarray}
\nonumber
\vec{v}_n(\eta) &\approx& c_n(\eta)\left[\alpha(\eta)\vec{\epsilon}_n^{\rm F}+(1-\alpha(\eta))\vec{\epsilon}_n^{\rm B}\right]\\\label{eq:6}
&&+\vec{\delta}_n^{\mathrm{geo}}(\eta)+\vec{\delta}_n^{\mathrm{dyn}}(\eta)\;,
\end{eqnarray}
where the $c_n(\eta)$ is the hydrodynamic response function, and the three additional terms in the form of $\vec{\delta}_n= \delta_ne^{in\sigma_n}$ represent additional initial or final state effects. The term $\vec{\delta}_n^{\mathrm{geo}}(\eta)$ represents additional geometric effects not accounted for by the eccentricity, such as the details in the radial distribution of the energy density profile~\cite{Teaney:2010vd,Floerchinger:2013vua} and the difference from an alternative definition of eccentricity~\cite{Qiu:2011iv}. The last term $\vec{\delta}_n^{\mathrm{dyn}}$ represents additional dynamical fluctuations~\cite{Gardim:2012im,Heinz:2013bua} generated during the hydrodynamic evolution and hadronization.

Secondly, $\vec{\epsilon}_n^{\rm F}$ and $\vec{\epsilon}_n^{\rm B}$ fluctuate strongly event to event, both in their magnitude and orientation. If $\epsilon_n^{\rm F}\neq\epsilon_n^{\rm B}$, the distributions of flow coefficients $v_n(\eta)$ are expected to show strong forward-backward asymmetry. Similarly, if $\Phi_n^{*\rm F}\neq\Phi_n^{*\rm B}$, the event-plane angle $\Phi_n$ is expected to rotate gradually from backward rapidity to the forward rapidity. However since $\alpha(\eta)$ is a non-linear function, these changes may also not be linear, especially when $\npartf$ and $\npartb$ values are very different such as in Cu+Au or p+Pb collisions.

\begin{figure}[t]
\begin{center}
\includegraphics[width=1\columnwidth]{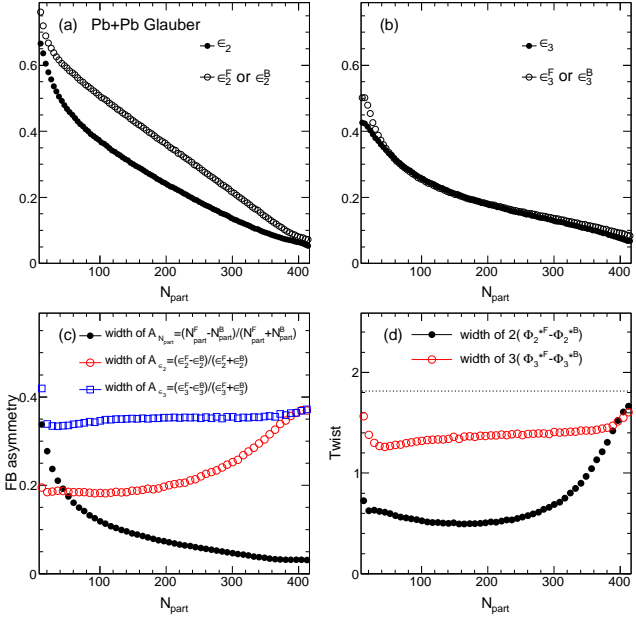}
\end{center}
\caption{\label{fig:ecc1} The ellipticity (panel (a)) and triangularity (panel (b)) calculated for all participating nucleons (filled circles) and using only forward-going participating nucleons (open circles). Panel (c) shows the RMS width of the FB asymmetry parameters defined in Eqs.~\ref{eq:3}-\ref{eq:3b} and panel (d) shows the RMS width of the twist angle between the eccentricity vectors of the two nuclei. The dotted-line indicate the value expected for flat twist distribution. All quantities are shown as a function of $\npart$.}
\end{figure}

A simple monte-carlo Glauber model~\cite{Miller:2007ri} is used to estimate the FB eccentricity fluctuations in Pb+Pb collisions. The results as a function of $\npart$ are summarized in Fig.~\ref{fig:ecc1}.
The values of $\epsilon_2^{\mathrm{F}}$ and $\epsilon_2^{\mathrm{B}}$ are found to be always larger than $\epsilon_2$ over the full centrality range (Figure~\ref{fig:ecc1}(a)), and this difference is due to the fact that the center-of-mass of the wounded nucleons in each nucleus is not at the center of the overlap region but is shifted towards the center of the corresponding nucleus (see discussion in Appendix~\ref{sec:app1}). In contrast, the values of $\epsilon_3^{\mathrm{F}}$ and $\epsilon_3^{\mathrm{B}}$ are similar to $\epsilon_3$.

The eccentricity vectors also exhibit a large FB asymmetry in their magnitude ($A_{\epsilon_n}$ in Figure~\ref{fig:ecc1} (c)) and a sizable twist (Figure~\ref{fig:ecc1} (d)). The asymmetry and the twist are nearly independent of centrality for $n=3$, but they are much smaller for $n=2$ in mid-central and peripheral collisions, reflecting the alignment of $\vec{\epsilon}_2^{\mathrm{F}}$ and $\vec{\epsilon}_2^{\mathrm{B}}$ to the almond shape of the overlap region.  In most central collisions, however, the width of the $A_{\epsilon_n}$ and twist angle for $n=2$ are comparable to that for $n=3$, reflecting a strong decorrelation between $\vec{\epsilon}_2^{\mathrm{F}}$ and $\vec{\epsilon}_2^{\mathrm{B}}$ due to the dominance of random fluctuations. According to Eq.~\ref{eq:6}, these FB asymmetry and twist should affect the longitudinal dynamics of harmonic flow. 

What we described so far are generic long-range initial state effects, which should be present as long as particle production associated with each wounded nucleon is not symmetric in the beam direction around the collision point. These effects are naturally included in any hydrodynamic models or transport models that includes realistic longitudinal dynamics. In the following, we describe a simulation analysis using the AMPT model~\cite{Lin:2004en}, and demonstrate that these initial state effects are indeed the main sources of longitudinal fluctuation of harmonic flow.
\section{simulation with the AMPT model} 
\label{sec:3}

The ``a multi-phase transport model'' (AMPT)~\cite{Lin:2004en} has been used frequently to study the higher-order $v_n$ associated with $\epsilon_n$~\cite{Xu:2011jm,Xu:2011fe,Ma:2010dv}. It combines the initial fluctuating geometry based on the Glauber model from HIJING and the final state interaction via a parton and hadron transport model, with the collective flow generated mainly by the partonic transport. The initial condition of the AMPT model contains significant longitudinal fluctuations that can influence the collective dynamics~\cite{Pang:2012he,Pang:2012uw,Huo:2013qma,Pang:2013pma}. The model simulation is performed with the string-melting mode with a total partonic cross-section of 1.5 mb and strong coupling constant of $\alpha_{\rm s}=0.33$~\cite{Xu:2011fe}. This setup has been shown to reproduce the experimental $\pT$ spectra and $v_n$ data at RHIC and the LHC~\cite{Xu:2011fe,Xu:2011fi}. 

\begin{figure}[t]
\centering
\includegraphics[width=1\linewidth]{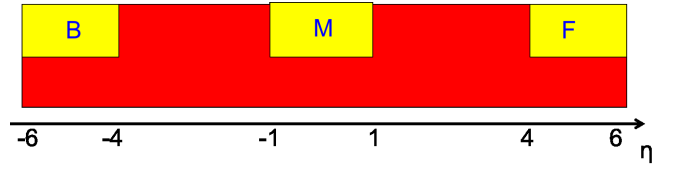}
\caption{\label{fig:1} The $\eta$-ranges of three subevents (B,M,F) for calculating flow vector $\vec{q}_n= q_ne^{in\Psi_n}$ via Eq.~\ref{eq:7}. They cover the pseudorapidity ranges of $-6<\eta<-4$ (backward or B), $-1<\eta<1$ (middle or M) and $4<\eta<6$ (forward or F). The results in this section (Sec.~\ref{sec:3}) are obtained using all particles in their respective $\eta$ ranges. But event planes used in Sec.~\ref{sec:5} are obtained using half of the particles, to allow the measurement of the flow harmonics over the full $\eta$ range.}
\end{figure}

The AMPT data used in this study is generated for $b=8$~fm Pb+Pb collisions at LHC energy of $\sqrt{s_{\mathrm{NN}}}=2.76$ TeV. A fraction of the particles in each event are divided into three subevents along $\eta$, $-6<\eta<-4$, $-1<\eta<1$ and $4<\eta<6$, as shown in Fig.~\ref{fig:1}, labelled as B, M and F. The raw flow vector in each subevent is calculated as:
\begin{eqnarray}
\label{eq:7}
\vec{q}_n  = q_ne^{in\Psi_n}=\frac{1}{\Sigma w}\left(\textstyle\Sigma (w\cos n\phi), i\Sigma (w\sin n\phi)\right)\;,
\end{eqnarray}
where the weight $w$ is chosen as the $\pT$ of each particle and $\Psi_n$ is the measured event-plane angle. Due to finite number effects, $\Psi_n$  smears around the true event-plane angle $\Phi_n$. If the FB eccentricity fluctuation is the source of rapidity fluctuation of harmonic flow, then we expect $\vec{q}_n^{\mathrm{F}}$ to be more correlated with $\vec{\epsilon}_n^{\mathrm{F}}$, $\vec{q}_n^{\mathrm{B}}$ to be more correlated with $\vec{\epsilon}_n^{\mathrm{B}}$ and $\vec{q}_n^{\mathrm{M}}$ to be more correlated with $\vec{\epsilon}_n$.

In each generated event, the following quantities are calculated for $n=2$ and 3: eccentricity quantities $\epsilon_n,\epsilon_n^{\mathrm{F}},\epsilon_n^{\mathrm{B}},\Phi_n^*,\Phi_n^{\mathrm{*F}},\Phi_n^{\mathrm{*B}}$, $(q_n^{\mathrm{F}}, \Psi^{\mathrm{F}}_n)$ for subevent F, $(q_n^{\mathrm{B}}, \Psi^{\mathrm{B}}_n)$ for subevent B and $(q_n^{\mathrm{M}}, \Psi^{\mathrm{M}}_n)$ for subevent M,  a total of 24 quantities. We also 
define the FB eccentricity difference, as well as  the twist angles between FB participant planes, the raw event planes and the true event planes as:
\begin{eqnarray}
\nonumber
\Delta\epsilon_n^{\mathrm{\mathrm{FB}}}&=&\epsilon_n^{\mathrm{F}}-\epsilon_n^{\mathrm{B}}\\\nonumber
\Delta\Phi_n^{\mathrm{\mathrm{*FB}}}&=&\Phi_n^{\mathrm{*F}}-\Phi_n^{\mathrm{*B}}\\\nonumber
\Delta\Psi_n^{\mathrm{\mathrm{FB}}}&=&\Psi_n^{\mathrm{F}}-\Psi_n^{\mathrm{B}}\\\label{eq:8}
\Delta\Phi_n^{\mathrm{\mathrm{FB}}}&=&\Phi_n^{\mathrm{F}}-\Phi_n^{\mathrm{B}}\;.
\end{eqnarray}
The twist angles are often represented as $n\Delta\Phi_n^{\mathrm{\mathrm{*FB}}}$, $n\Delta\Psi_n^{\mathrm{\mathrm{FB}}}$ or $n\Delta\Phi_n^{\mathrm{\mathrm{FB}}}$ such that their periods are always $2\pi$.

\begin{figure}[h]
\begin{center}
\includegraphics[width=1\columnwidth]{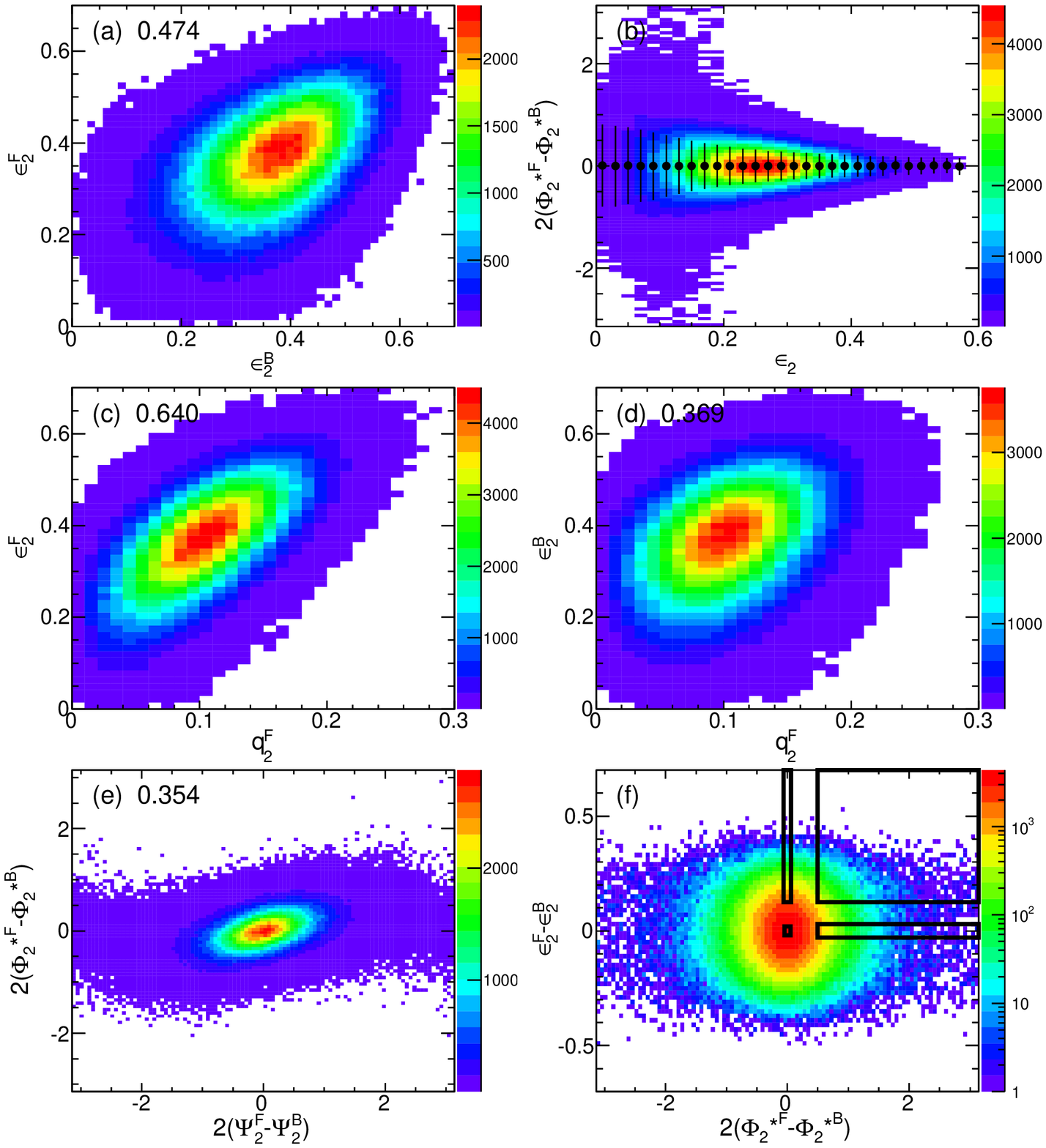}
\end{center}
\caption{\label{fig:2} Correlation of $\epsilon_2^{\mathrm{B}}$ vs. $\epsilon_2^{\mathrm{F}}$ (a), $2\Delta\Phi_2^{\mathrm{*FB}}$ vs. $\epsilon_2$ (b),  $\epsilon_2^{\mathrm{F}}$ vs. $q_2^{\mathrm{F}}$ (c),  $\epsilon_2^{\mathrm{F}}$ vs. $q_2^{\mathrm{B}}$ (d),  $2\Delta\Phi_2^{\mathrm{*FB}}$ vs. $2\Delta\Psi_2^{\mathrm{FB}}$ (e), and $\epsilon_2^{\mathrm{F}}-\epsilon_2^{\mathrm{F}}$ vs. $2\Delta\Phi_2^{\mathrm{*FB}}$ (f) for AMPT Pb+Pb events with $b=8$~fm. The bars around the circles in panel-(b) indicates the RMS width of $2\Delta\Phi_2^{\mathrm{*FB}}$ at given $\epsilon_2$ value, and the four regions delineated by the boxes in panel-(f) indicate the cuts for the four event classes defined in Table~\ref{tab:cut}. The numbers in some panels indicates the correlation coefficients of the distributions.}
\end{figure}

Figure~\ref{fig:2} summarizes the FB correlations between eccentricity vectors and the flow vectors, together with the corresponding correlation coefficients. A positive correlation is observed between $\epsilon_2^{\mathrm{F}}$ and $\epsilon_2^{\mathrm{B}}$ in Fig.~\ref{fig:2}(a), and they both are also positively correlated with $\epsilon_2$ (not shown). The PP angles $\Phi_2^{\mathrm{*F}}$ and $\Phi_2^{\mathrm{*B}}$ are not perfectly aligned (Fig.~\ref{fig:2}(b)). The width of $2(\Phi_2^{\mathrm{*F}}-\Phi_2^{\mathrm{*B}})$ is about 0.21 radian for top 10\% of events with largest $\epsilon_2$ and increases to 0.70 radian for the bottom 10\% of events with smallest $\epsilon_2$. Thus significant FB asymmetry and twist in the ellipticity are expected for most of the events. 

Figures~\ref{fig:2}(c) and \ref{fig:2}(d) show the correlation of $\epsilon_2^{\mathrm{F}}$ with $q_2^{\mathrm{F}}$ and $q_2^{\mathrm{B}}$, respectively. The correlation is stronger between $\epsilon_2^{\mathrm{F}}$ and $q_2^{\mathrm{F}}$ than that between $\epsilon_2^{\mathrm{F}}$ and $q_2^{\mathrm{B}}$, suggesting that the elliptic flow in the forward-rapidity is driven more by the ellipticity of the forward-going Pb nucleus (and vice versa). This is expected since the forward particle production arises preferably from forward-going participating nucleons. Figure~\ref{fig:2}(e) shows that the angles between the participant planes are strongly correlated with the raw event planes, suggesting that the twist in the initial state geometry is converted into twist in the final collective flow between the forward and the backward pseudorapidity.

\begin{figure}[h]
\begin{center}
\includegraphics[width=1\columnwidth]{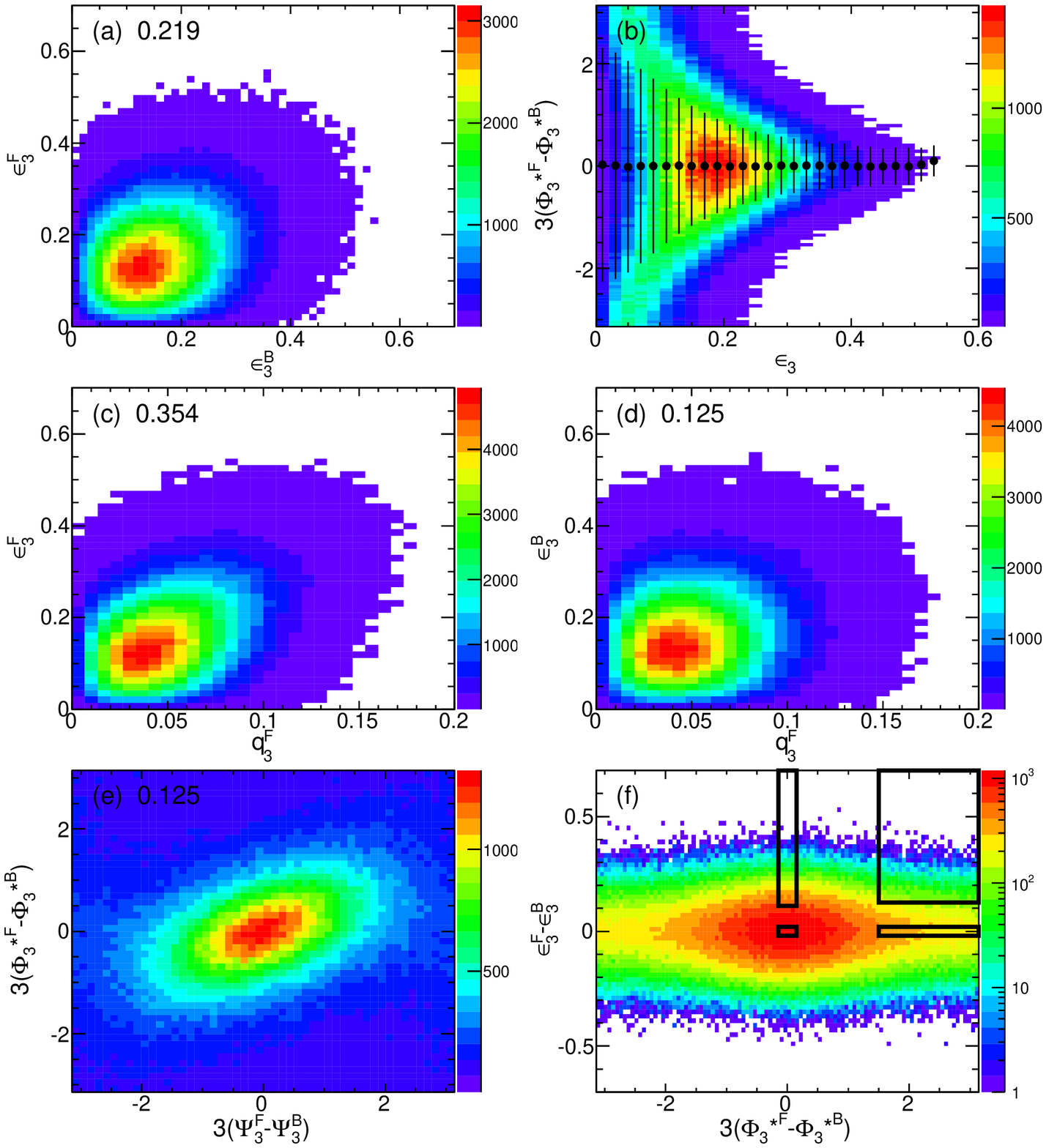}
\end{center}
\caption{\label{fig:3}Correlation of $\epsilon_3^{\mathrm{B}}$ vs. $\epsilon_3^{\mathrm{F}}$ (a), $3\Delta\Phi_3^{\mathrm{*FB}}$ vs. $\epsilon_3$ (b),  $\epsilon_3^{\mathrm{F}}$ vs. $q_3^{\mathrm{F}}$ (c),  $\epsilon_3^{\mathrm{F}}$ vs. $q_3^{\mathrm{B}}$ (d),  $3\Delta\Phi_3^{\mathrm{*FB}}$ vs. $3\Delta\Psi_3^{\mathrm{FB}}$ (e), and $\epsilon_3^{\mathrm{F}}-\epsilon_3^{\mathrm{F}}$ vs. $3\Delta\Phi_3^{\mathrm{*FB}}$ (f) for AMPT Pb+Pb events with $b=8$~fm. The bars around the circles in panel-(b) indicates the RMS width of $3\Delta\Phi_3^{\mathrm{*FB}}$ at given $\epsilon_3$ value, and the four regions delineated by the boxes in panel-(f) indicate the cuts for the four event classes defined in Table~\ref{tab:cut}. The numbers in some panels indicates the correlation coefficients of the distributions.}
\end{figure}

Identical studies are also performed for the triangularity and triangular flow in Fig.~\ref{fig:3}.
The features are qualitatively similar to those shown in Fig.~\ref{fig:2}, except that most forward-backward correlations are significantly weaker, as $\vec{\epsilon}_3^{\mathrm{F}}$ and $\vec{\epsilon}_3^{\mathrm{F}}$ are both dominated by random fluctuations. In particular, the distribution of twist angle $3(\Phi_3^{\mathrm{\mathrm{*F}}}-\Phi_3^{\mathrm{\mathrm{*B}}})$ is much broader than that of $2(\Phi_2^{\mathrm{\mathrm{*F}}}-\Phi_2^{\mathrm{\mathrm{*B}}})$ in Fig.~\ref{fig:2}(b). In fact, $\Phi_3^{\mathrm{\mathrm{*F}}}$ and $\Phi_3^{\mathrm{\mathrm{*B}}}$ are nearly out-of-phase for events selected with small $\epsilon_3$ (the lower 30\% of events). This large twist is the dominating source of the decorrelation of triangular flow observed in previous studies~\cite{Xiao:2012uw,Jia:2014vja}. 

Given the rich patterns of the FB eccentricity and PP-angle fluctuations shown in Figs.~\ref{fig:2} and \ref{fig:3}, the plan of this paper is not to perform an exhaustive study of the collective response of all possible configurations of the initial geometry. Instead, we focus on several representative classes of events and study how their specific initial state configurations influence the $\vec{v}_n(\eta)$ values in the final state. Four event classes are defined separately for ellipticity and triangularity in Table~\ref{tab:cut} by cutting on $\Delta\epsilon_n^{\mathrm{*FB}}$ and $n\Delta\Phi_n^{\mathrm{*FB}}$, they are also indicated visually in Fig.~\ref{fig:2} (f) for $n=2$ and Fig.~\ref{fig:3}(f) for $n=3$. The ``type1'' events have nearly identical initial shape between the two nuclei, i.e.  $(\epsilon_n^{\mathrm F},\Phi_n^{\mathrm{*F}})\approx (\epsilon_n^{\mathrm B},\Phi_n^{\mathrm{*B}})$. The ``type2'' events have similar PP angles but very asymmetric eccentricity values, i.e $\Phi_n^{\mathrm{*F}}\approx\Phi_n^{\mathrm{*B}}$ and $\epsilon_n^{\mathrm F}>\epsilon_n^{\mathrm B}$. The ``type3'' events have similar eccentricity values but large twist between the two nuclei, i.e. $\epsilon_n^{\mathrm F}\approx\epsilon_n^{\mathrm B}$ and $\Phi_n^{\mathrm{*F}}>\Phi_n^{\mathrm{*B}}$. The ``type4'' events have large twist angle as well as very asymmetric eccentricity values. Each class contains at least 1.5\% of the total event statistics, so they represent some typical events with very different initial condition. 

\begin{table}[h]
\centering
\begin{tabular}{|c|c|c|c|}\hline
\multicolumn{4}{|c|}{event classes in ellipticity}\\\hline
    &Cuts                  & $\langle\epsilon_2^{\rm F}\rangle$ & $\langle\epsilon_2^{\rm B}\rangle$\\\hline
type1& $|2\Delta\Phi_2^{\mathrm{*FB}}|<0.05,|\Delta\epsilon_2^{\mathrm{FB}}|<0.02$ &\multicolumn{2}{|c|}{0.4}\\\hline
type2& $|2\Delta\Phi_2^{\mathrm{*FB}}|<0.05,\Delta\epsilon_2^{\mathrm{FB}}>0.125$  &0.456&0.282\\\hline
type3& $2\Delta\Phi_2^{\mathrm{*FB}}>0.5,|\Delta\epsilon_2^{\mathrm{FB}}|<0.03$    &\multicolumn{2}{|c|}{0.314}\\\hline
type4& $2\Delta\Phi_2^{\mathrm{*FB}}>0.5,\Delta\epsilon_2^{\mathrm{FB}}>0.125$     &0.386&0.197 \\\hline\hline
\multicolumn{4}{|c|}{event classes in triangularity}\\\hline
 &Cuts                    & $\langle\epsilon_3^{\rm B}\rangle$   & $\langle\epsilon_3^{\rm B}\rangle$\\\hline
type1& $|3\Delta\Phi_3^{\mathrm{*FB}}|<0.15,|\Delta\epsilon_3^{\mathrm{FB}}|<0.02$ &\multicolumn{2}{|c|}{0.182}\\\hline
type2& $|3\Delta\Phi_3^{\mathrm{*FB}}|<0.15,\Delta\epsilon_3^{\mathrm{FB}}>0.125$  &0.293&0.126\\\hline
type3& $3\Delta\Phi_3^{\mathrm{*FB}}>1.5,|\Delta\epsilon_3^{\mathrm{FB}}|<0.02$    &\multicolumn{2}{|c|}{0.112}\\\hline
type4& $3\Delta\Phi_3^{\mathrm{*FB}}>1.5,\Delta\epsilon_3^{\mathrm{FB}}>0.125$     &0.246&0.0686\\\hline\hline
\end{tabular}
\caption{\label{tab:cut} The four classes of events selected on $n\Delta\Phi_n^{\mathrm{*FB}}=n(\Phi_n^{\mathrm{*F}}-\Phi_n^{\mathrm{*B}})$ and $\Delta\epsilon_n^{\mathrm{FB}}=\epsilon_n^{\rm F}-\epsilon_n^{\rm B}$ for $n=2$ (top half) and $n=3$ (bottom half). The corresponding average eccentricity values for the two colliding Pb nuclei are also listed.}
\end{table}

In order to study the rapidity fluctuation of harmonic flow, we need to calculate the Fourier coefficients for final state particles relative to a $n^{\mathrm{th}}$-order reference plane angle $\Theta_n$ in each event:
\begin{eqnarray}
\nonumber
\vnc (\eta) &=& \langle\cos n\left(\phi(\eta)-\Theta_n\right)\rangle\\\nonumber
\vns (\eta) &=& \langle\sin n\left(\phi(\eta)-\Theta_n\right)\rangle\\\nonumber
v_n(\eta)&=& \sqrt{(\vnc(\eta))^2+(\vns(\eta))^2}\\\label{eq:9}
\tan \left[n\Delta\Phi_n^{\mathrm{rot}}(\eta)\right] &=&\frac{\langle\sin n\left(\phi(\eta)-\Theta_n\right)\rangle}{\langle\cos n\left(\phi(\eta)-\Theta_n\right)\rangle}= \frac{\vns(\eta)}{\vnc(\eta)}\;.
\end{eqnarray}
where the average is over all particles at $\eta$ in the event, and $n\Delta\Phi_n^{\mathrm{rot}}(\eta)$ is the $\eta$-dependent twist of the final state particles relative to $\Theta_n$. The angle $\Theta_n$ is chosen as one of the participant-plane angles $\Theta_n\in\{\Phi_n^{\mathrm{*F}},\Phi_n^{\mathrm{*B}}, \Phi_n^{\mathrm{*}}\}$, or the truth event-plane angle in one of three subevents, $\Theta_n\in\{\Phi_n^{\mathrm{F}},\Phi_n^{\mathrm{B}}, \Phi_n^{\rm M}\}$. In general, the Fourier coefficients may not be the same between the participant plane and event plane, especially if $\Phi_n^{\mathrm{*F}}\neq\Phi_n^{\mathrm{*B}}$ or the $\vec{\delta}_n$ terms in Eq.~\ref{eq:6} are important. 

The quantities in Eq.~\ref{eq:9} in principle contains the complete information of the harmonic flow in one event, and they can be calculated precisely in an EbyE full 3+1D hydrodynamic simulation. But the calculation is not possible in the actual experiment since the number of particles in one event is finite. Instead, the averages in Eq.~\ref{eq:9} have to be performed over an ensemble, which is typically chosen as all events in one centrality class in previous flow analyses. But since the probability of events with positive twist and negative twist are identical, the sine component always cancels out, $\vns (\eta)=0$, while the cosine component $\vnc$ is unaffected. Hence the results averaged over events in a centrality class always underestimate the true flow amplitudes, and the twist leads to a decrease of measured $v_n=\vnc$ in $\eta$ away from the reference plane (first observed in Ref.~\cite{Jia:2014vja}, but see also Figure~\ref{fig:res4b}).


To directly expose the influence of the twist, events in an ensemble must be selected to preferably rotate in one direction, such as the ``type3'' and ``type4'' events in Table~\ref{tab:cut}. For these ``helicity''-selected ensembles, the event-averaged twist angle as a function of $\eta$ can be measured with Eq.\ref{eq:9}, where the average is now performed over all particles at $\eta$ in one event, then over all events in the ensemble. We emphasize that there could be other sources of EbyE twist associated with the $\vec{\delta}_n$ terms in Eq.\ref{eq:6}. If these sources are orthogonal to event selection criteria, i.e. probabilities for positive twist and negative twist from these sources are the same, they would not contribute to $\vns (\eta)$ after event-average, and hence the measurements generally underestimate the true flow signal.

In order to obtain the Fourier coefficients associated with the true event plane, the $v_n^{\mathrm{c,raw}}$ and $v_n^{\mathrm{s,raw}}$ measured relative to raw event plane need to be corrected by a resolution factor determined via a standard method~\cite{Aad:2012bu} (with the caveat discussed in Appendix~\ref{sec:app2}):
\begin{eqnarray}
\label{eq:11}
v_n^{\mathrm{c}}(\eta)= \frac{v_n^{\mathrm{c,raw}}(\eta)}{\mathrm{Res}\{ n\Psi_n \} }\;,\;\;\; v_n^{\mathrm{s}}(\eta)= \frac{v_n^{\mathrm{s,raw}}(\eta)}{\mathrm{Res}\{ n\Psi_n \} }\;.
\end{eqnarray}
This correction changes the sine and cosine coefficients by the same amount, hence the twist angle relative to the event plane can be obtained directly from the raw Fourier coefficients as ~\cite{Jia:2014vja}:
\begin{eqnarray}
\label{eq:12}\tan \left[n\Delta\Phi_n^{\mathrm{rot}}(\eta)\right]=\frac{\vnsr}{\vncr}\;.
\end{eqnarray}

Due to the presence of the $\vec{\delta}_n^{\mathrm{geo}}(\eta)$ and $\vec{\delta}_n^{\mathrm{dyn}}$ terms in Eq.~\ref{eq:6}, the participant-plane angles may also fluctuate randomly around the true event plane (observed in EbyE hydrodynamics calculations~\cite{Qiu:2011iv,Niemi:2012aj}), and correction factors similar to Eq.~\ref{eq:11} are also needed. These corrections are assumed to be small and are neglected in this paper.

In the following section, we first discuss the $v_n(\eta)$ and $n\Delta\Phi_n^{\mathrm{rot}}(\eta)$ results obtained from the three participant planes $\Phi_n^{\mathrm{*F}},\Phi_n^{\mathrm{*B}}$, and $\Phi_n^{\mathrm{*}}$ in each event class. These results provide clear understanding on how the flow harmonics after collective response in AMPT correlates with specific choices of initial geometry characterized by $\vec{\epsilon}_n^{\rm F}$, $\vec{\epsilon}_n^{\rm B}$ and $\vec{\epsilon}_n$. We then discuss results obtained using the three event planes $\Psi_n^{\mathrm{F}},\Psi_n^{\mathrm{B}}$, and $\Psi_n^{\rm M}$ and compare with the participant plane results. This comparison allow us to quantify how well the calculations based on the experimental method reproduce the true correlation with the initial geometry.

\section{Results based on participant planes} 
\label{sec:4}

\begin{figure}[t]
\begin{center}
\includegraphics[width=1\columnwidth]{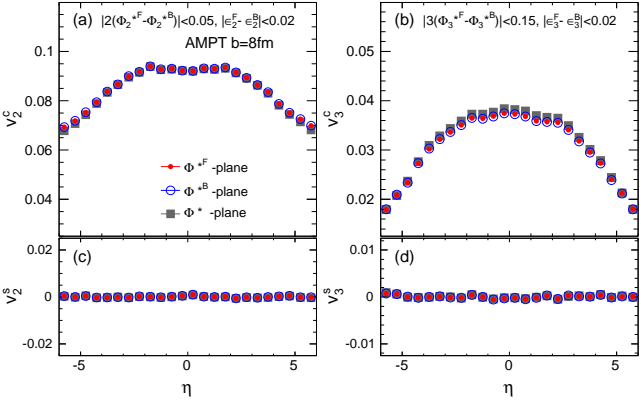}
\end{center}
\caption{\label{fig:res1} The $\vnc(\eta)$ (top row) and $\vns(\eta)$ (second row) relative to the reference angle taken as one of the three participant planes. They are obtained for ``type1'' events for $n=2$ (left column) and $n=3$ (right column).}
\end{figure}

Figure~\ref{fig:res1} shows the elliptic (left) and triangular (right) flow harmonics for events selected with the ``type1'' criteria in Table~\ref{tab:cut}. The values of $\vtwos$ and $\vthrs$ are consistent with zero. The values of $\vtwoc$ and $\vthrc$ are symmetric around $\eta=0$, and they are nearly identical between the three participant planes. These behaviors are expected since the initial geometry of the ``type1'' events are symmetric between the forward-going and backward-going nuclei.

\begin{figure}[t]
\begin{center}
\includegraphics[width=1\columnwidth]{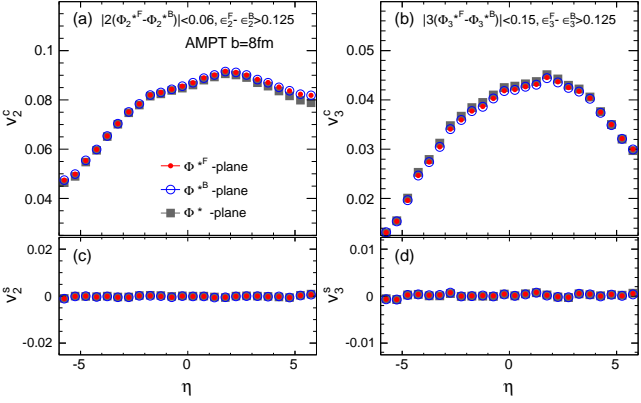}
\end{center}
\caption{\label{fig:res2} The $\vnc(\eta)$ (top row) and  $\vns(\eta)$ (second row) relative to the reference angle taken as one of the three participant planes. They are obtained for ``type2'' events for $n=2$ (left column) and $n=3$ (right column).}
\end{figure}

Figure~\ref{fig:res2} shows the flow harmonics for events selected with the ``type2'' criteria in Table~\ref{tab:cut}. The values of $\vtwos$ and $\vthrs$ are consistent with zero, reflecting the requirement that $\Phi_n^{\mathrm{*F}}\approx\Phi_n^{\mathrm{*B}}$. However the values of $\vtwoc$ and $\vthrc$ are significantly larger in the forward $\eta$ than the backward $\eta$. This apparent asymmetry is attributed to the requirement that $\epsilon_n^{\mathrm{F}}>\epsilon_n^{\mathrm{B}}$.

\begin{figure}[t]
\begin{center}
\includegraphics[width=1\columnwidth]{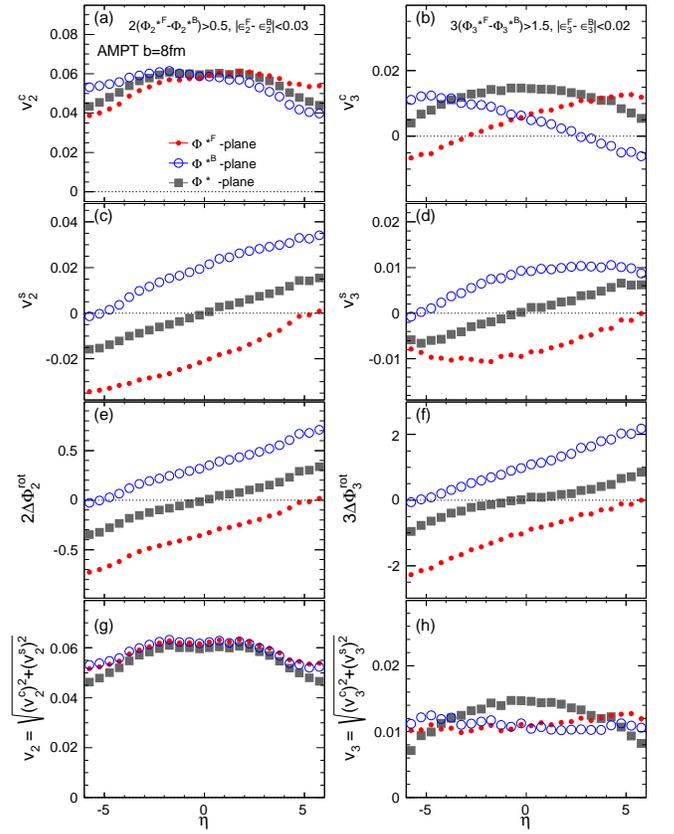}
\end{center}
\caption{\label{fig:res3} The $\vnc(\eta)$ (top row), $\vns(\eta)$ (second row), rotation angle $n\Delta\Phi_n^{\mathrm{rot}}$ (third row) and $v_n=\sqrt{(\vnc)^2+(\vns)^2}$ (bottom row) relative to the reference angle taken as one of the three participant planes. They are obtained via Eq.~\ref{eq:12} for ``type3'' events for $n=2$ (left column) and $n=3$ (right column).}
\end{figure}

Figure~\ref{fig:res3} shows the flow harmonics for events selected with the ``type3'' criteria in Table~\ref{tab:cut}. The FB eccentricity vectors in these events have similar magnitudes, but are twisted relative to each other. This initial twist leads to significant nonzero $\vns(\eta)$ values. The extracted twist angle $n\Delta\Phi_n^{\mathrm{rot}}(\eta)$ is found to vary linearly from the backward to the forward pseudorapidity. Significant FB asymmetry is also observed for $\vnc(\eta)$ calculated relative to $\Phi_n^{\mathrm{*F}}$ and $\Phi_n^{\mathrm{*B}}$. This asymmetry is especially large for triangular flow, reflected by an apparent sign flip at the two ends of the $\eta$ range. The truth flow magnitudes obtained by including the sine component, $v_n=\sqrt{(\vnc)^2+(\vns)^2}$, are nearly symmetric in $\eta$, consistent with the condition $\epsilon_n^{\mathrm{F}}\approx\epsilon_n^{\mathrm{B}}$.

Figure~\ref{fig:res3} also shows that the $v_3(\eta)$ values depend strongly on the choice of the reference plane: the values are largest for $\Phi_3^{\mathrm{*F}}$ at $\eta>4$, $\Phi_3^{*}$ at $\eta\approx 0$, and $\Phi_3^{\mathrm{*B}}$ at $\eta<-4$. This is because $\Phi_3^{\mathrm{*tot}}(\eta)$ is close to $\Phi_3^{\mathrm{*F}}$ for $\eta>4$, close to $\Phi_3^{*}$ for $\eta\approx 0$, and close to $\Phi_3^{\mathrm{*B}}$ for $\eta<-4$. These behaviors suggest that when the twist angles are large (they are nearly out-of-phase for $n=3$), the three participant planes are only a good approximation of $\Phi_3^{\mathrm{*tot}}(\eta)$ of Eq.~\ref{eq:6} over a limited $\eta$ range.

\begin{figure}[t]
\begin{center}
\includegraphics[width=1\columnwidth]{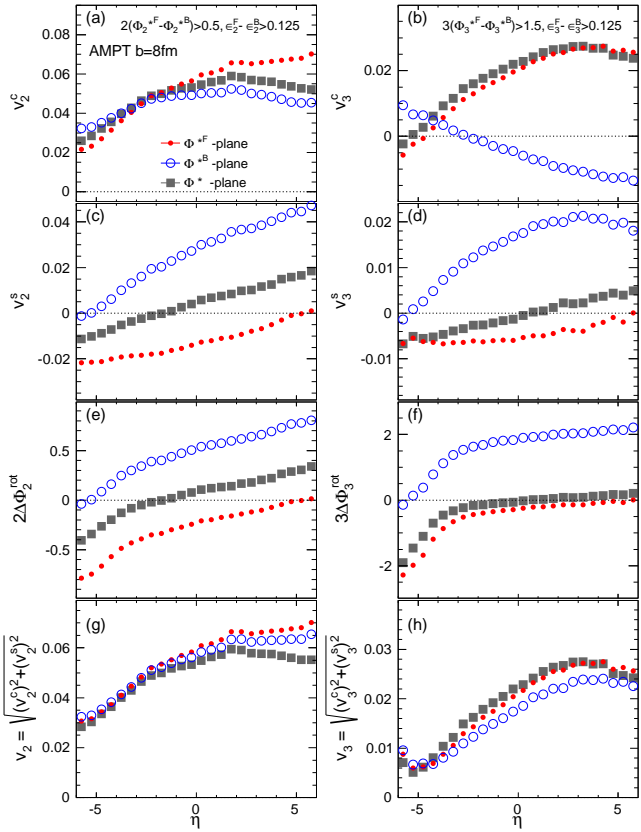}
\end{center}
\caption{\label{fig:res4} The $\vnc(\eta)$ (top row), $\vns(\eta)$ (second row), rotation angle $n\Delta\Phi_n^{\mathrm{rot}}$ (third row) and $v_n=\sqrt{(\vnc)^2+(\vns)^2}$ (bottom row) relative to the reference angle taken as one of the three participant planes. They are obtained via Eq.~\ref{eq:12} for ``type4'' events for $n=2$ (left column) and $n=3$ (right column).}
\end{figure}

Figure~\ref{fig:res4} shows the flow harmonics for events selected with the ``type4'' criteria in Table~\ref{tab:cut}. The flow coefficients have similar properties with those in Fig.~\ref{fig:res3}, except that they have strong FB asymmetries due to $\epsilon_n^{\mathrm{F}}\gg\epsilon_n^{\mathrm{B}}$. For the same reason, the flow coefficients and the twist angles for $\Phi_n^{*}$ are similar with those for $\Phi_n^{\mathrm{*F}}$. The influence of $\vec{\epsilon}_n^{\mathrm{B}}$ is significant only in the very backward rapidity ($\eta<-4$), reflected by the sharp drop of $\Delta\Phi_n^{\mathrm{rot}}(\eta)$ in Fig.~\ref{fig:res4}(f). 

\begin{figure}[h]
\begin{center}
\includegraphics[width=1\columnwidth]{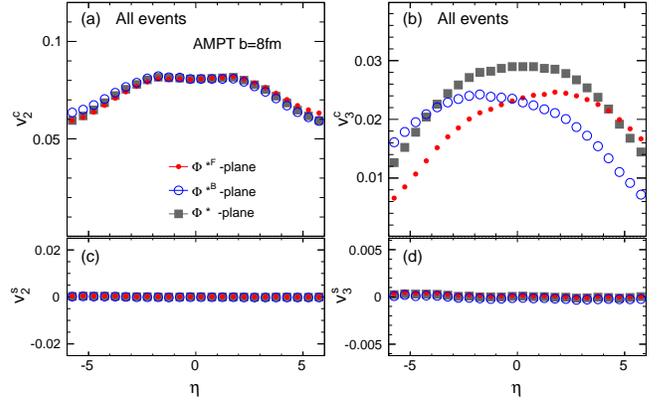}
\end{center}
\caption{\label{fig:res4b} The $\vnc(\eta)$ (top row) and $\vns(\eta)$ (second row) relative to the reference angle taken as one of the three participant planes for $n=2$ (left column) and $n=3$ (right column). No selection cuts have been applied for these events.}
\end{figure}

Once we understand the behavior of $v_n$ in the four event classes discussed above, it is straightforward to discuss the influence of the FB eccentricity and PP angle fluctuations on the flow harmonics for all events without any selection cuts, as shown in Fig.~\ref{fig:res4b}. The $\vns$ values vanish since the probabilities for positive and negative twist are the same. The $\vnc$ values at given $\eta$ show a characteristic hierarchy between the results for the three participant planes: they are largest for $\Phi_n^{\mathrm{*F}}$ in the forward rapidity, for $\Phi_n^{\mathrm{*}}$ in the mid-rapidity, and for $\Phi_n^{\mathrm{*B}}$ in the backward-rapidity, respectively. For triangular flow, the strong $\eta$-dependence and large spread between the results for the three participant planes are due to the large EbyE twist between $\vec{\epsilon}_3^{\mathrm{F}}$ and $\vec{\epsilon}_3^{\mathrm{B}}$ (see Figs.~\ref{fig:ecc1} and \ref{fig:3}). Similar hierarchy is also seen for elliptic flow but the differences are much smaller. If these decorrelation effects are important in the data, one would expect the $v_3$ results measured relative to the forward event plane to differ significantly from those measured relative to the backward event plane. Previous experimental analyses~\cite{Adare:2011tg,Adamczyk:2013waa,ALICE:2011ab,Aad:2012bu,Chatrchyan:2013kba,Aad:2013xma,Aad:2014fla} haven not observed such effects possibly because of the use of $\eta$-symmetric event planes.

Figure~\ref{fig:res4b}(a) also shows a small but visible double peak structure at $\eta\approx\pm2$ in the $\vtwoc(\eta)$ distributions. This feature is simply due to $\epsilon_2^{\mathrm{F/B}}>\epsilon_2$ (see Fig.~\ref{fig:ecc1} (a)), which slightly pushes up the $\vtwoc(\eta)$ at $\eta\pm2$ where the emission function $f^{\mathrm{F/B}}(\eta)$ (Eq.~\ref{eq:5b} and Fig.~\ref{fig:idea}) reaches maximum. Because of this, $\vtwoc(\eta)$ distribution is expected to be slightly broader than the $\vthrc(\eta)$ distribution, a feature we also observe in the LHC data~\cite{Aad:2012bu,Chatrchyan:2013kba}.

\section{Comparison with event plane results} 
\label{sec:5}

\begin{figure}[h]
\begin{center}
\includegraphics[width=1\columnwidth]{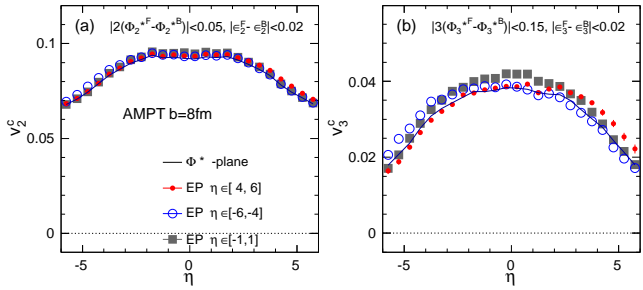}
\end{center}
\caption{\label{fig:res5} The $\vnc(\eta)$ obtained with the three raw event planes (see Fig.~\ref{fig:1}) for the ``type1'' events. They are compared with those obtained from the participant plane for all wounded nucleons, $\Phi_n^{*}$, from Fig.~\ref{fig:res1}.}
\end{figure}

\begin{figure}[h]
\begin{center}
\includegraphics[width=1\columnwidth]{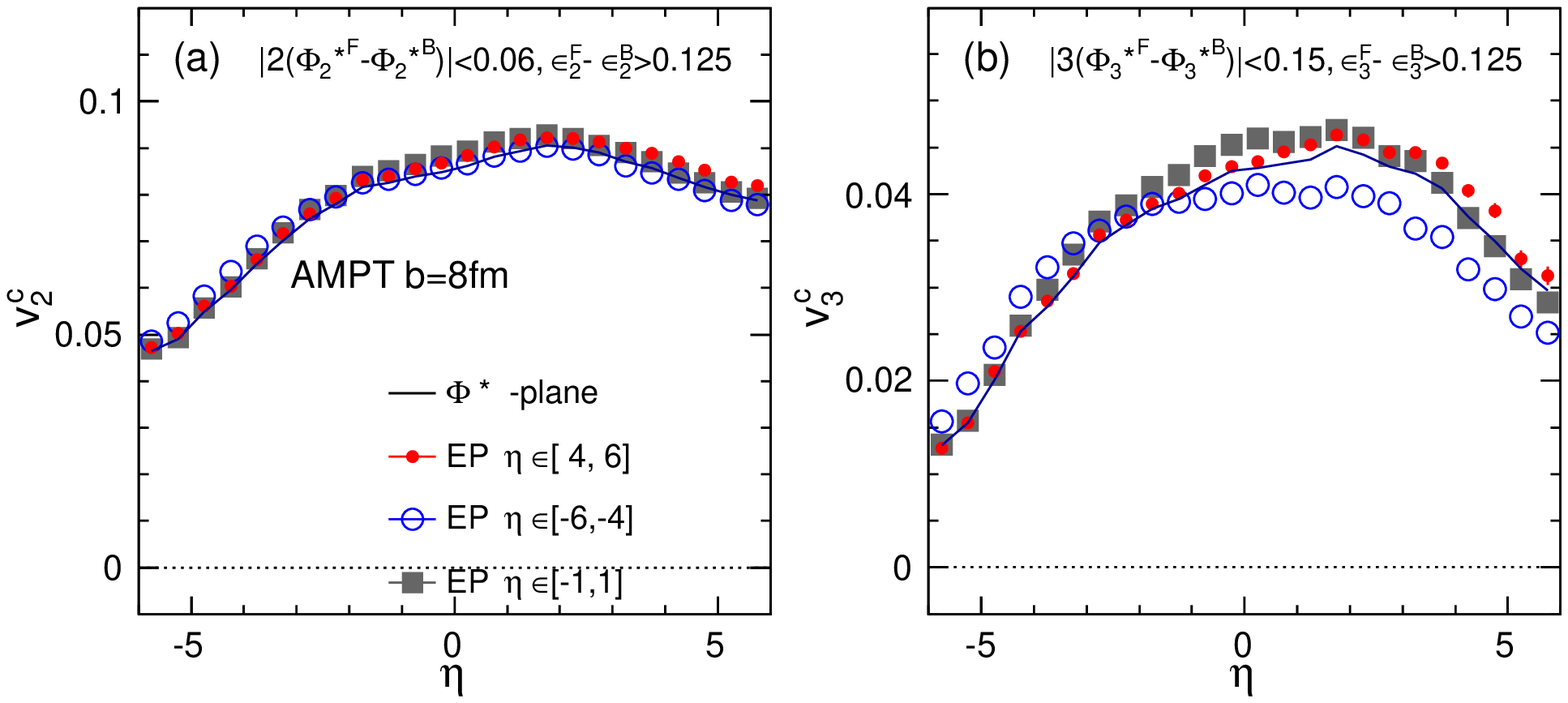}
\end{center}
\caption{\label{fig:res6} The $\vnc(\eta)$ obtained with the three EPs for the ``type2'' events. They are compared with those obtained from the participant plane for all wounded nucleons, $\Phi_n^{*}$, from Fig.~\ref{fig:res2}.}
\end{figure}

Figures~\ref{fig:res5} and \ref{fig:res6} show the flow harmonics for ``type1'' and ``type2'' events, calculated with the three raw event planes $\Psi_n^{\mathrm{B}},\Psi_n^{\rm M}$ and $\Psi_n^{\mathrm{F}}$ defined in Fig.~\ref{fig:1}. The results are compared with those obtained with $\Phi_n^{*}$ in Figs.~\ref{fig:res1} and~\ref{fig:res2}. The EP results quantitatively agree with the PP results in most cases, including the FB-asymmetry for ``type2'' events. Small systematic deviations are observed for $\vnc$ in $\eta$ region where the event planes are defined, reflecting a modest contribution from non-flow effects. 

\begin{figure}[h]
\begin{center}
\includegraphics[width=1\columnwidth]{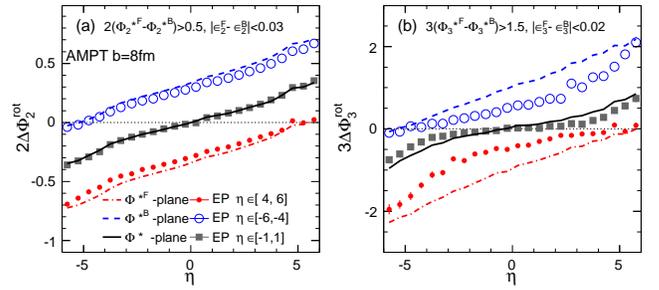}
\end{center}
\caption{\label{fig:res7} The twist angles obtained with the three EPs for the ``type3'' events. They are compared with those obtained from the three participant plane for all wounded nucleons, $\Phi_n^{*}$, from Fig.~\ref{fig:res3}.}
\end{figure}
\begin{figure}[h]
\begin{center}
\includegraphics[width=1\columnwidth]{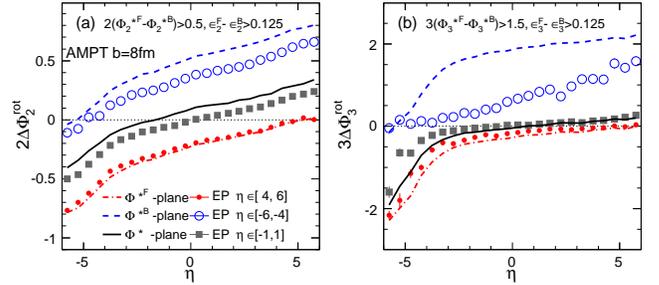}
\end{center}
\caption{\label{fig:res8}  The twist angles obtained with the three EPs for the ``type4'' events. They are compared with those obtained from the three participant plane for all wounded nucleons, $\Phi_n^{*}$, from Fig.~\ref{fig:res4}.}
\end{figure}

Similarly we also calculate the flow harmonics obtained from the event planes and compare them with the PP results for type3 and type4 events. Figures~\ref{fig:res7} and \ref{fig:res8} show that the twist angles for $\Psi_n^{\mathrm{F}}$,$\Psi_n^{\mathrm{B}}$ and $\Psi_n^{\mathrm{M}}$ approximately match those for $\Phi_n^{\mathrm{*F}}$, $\Phi_n^{\mathrm{*B}}$ and $\Phi_n^{\mathrm{*}}$, respectively.
However, several noticeable exceptions are observed for  $3\Delta\Phi_3^{\mathrm{rot}}(\eta)$ calculated with $\Psi_3^{\mathrm{F}}$ and $\Psi_3^{\mathrm{B}}$. These exceptions can be understood based on Eqs.~\ref{eq:5}-\ref{eq:6}: There are significant but unequal mixing between $\Phi_n^{\mathrm{*F}}$ and $\Phi_n^{\mathrm{*B}}$ in the $4<|\eta|<6$ where $\Psi_n^{\mathrm{F}}$ and $\Psi_n^{\mathrm{B}}$ are calculated. The raw EP angles $\Psi_3^{\mathrm{F}}$ and $\Psi_3^{\mathrm{B}}$ hence reflect the detailed interplay between the $\alpha(\eta)\vec{\epsilon}_3^{\mathrm{F}}$ term and the $(1-\alpha(\eta))\vec{\epsilon}_3^{\mathrm{B}}$ term in Eq.~\ref{eq:5}. 

For example, due to the dominance of  $\vec{\epsilon}_3^{\mathrm{F}}$ implied by the condition $\epsilon_3^{\mathrm{F}}>>\epsilon_3^{\mathrm{B}}$ for ``type4'' events, the values of $3\Delta\Phi_3^{\mathrm{rot}}(\eta)$ should be similar between $\Psi_3^{\mathrm{F}}$ and $\Psi_3^{\mathrm{M}}$. On the other hand, the $\Psi_3^{\mathrm{B}}$ is controlled by both $\vec{\epsilon}_3^{\mathrm{F}}$ and $\vec{\epsilon}_3^{\mathrm{B}}$, and hence the angle $\Psi_3^{\mathrm{B}}$ can different significantly from $\Phi_3^{\mathrm{*B}}$ as shown in Fig.~\ref{fig:res8}(b).

\section{Discussion and conclusion} 
\label{sec:6}

It is now commonly believed that the harmonic flow in heavy-ion collisions are the result of hydrodynamic response to the spatial fluctuations of the transverse density profile in the initial state. These spatial fluctuations are often obtained from a Glauber model~\cite{Miller:2007ri} or from a more advanced implementation that consider the quantum fluctuations of the participating nucleons and their sub-nucleonic structures~\cite{Schenke:2012fw}. The eccentricity vectors $\vec{\epsilon}_n=\epsilon_ne^{in\Phi_n^*}$ has been assumed to be applicable to a wide rapidity range (boost invariant). 

In this paper, we argue that since the shape of the participants in the two colliding nuclei can fluctuate independently, and that the energy deposition of each wounded nucleon is biased in $\eta$ along its direction of motion, the shape of the produced fireball at early time should be a strong function of $\eta$. The resulting eccentricity vector, which seeds the hydrodynamic evolution, should also be a strong function of $\eta$ in both the magnitude and the phase $\vec{\epsilon}^{\mathrm{tot}}_n(\eta)=\epsilon^{\mathrm{tot}}_n(\eta)e^{in\Phi_n^{\mathrm*tot}(\eta)}$. The $\vec{\epsilon}^{\mathrm{tot}}_n(\eta)$ should approximately interpolate between the eccentricity vectors calculated separately for the two colliding nuclei: $\vec{\epsilon}_n^{\rm F}$ in the far forward rapidity and $\vec{\epsilon}_n^{\rm B}$ in the far backward rapidity. The vector $\vec{\epsilon}_n$ is a good approximation for $\vec{\epsilon}^{\mathrm{tot}}_n(\eta)$ only at around the mid-rapidity. Large event-by-event fluctuations of $\vec{\epsilon}_n^{\rm F}$ and $\vec{\epsilon}_n^{\rm B}$ naturally produce a fireball that is asymmetric in their shape (i.e. $\epsilon_n^{\rm F}\neq\epsilon_n^{\rm B}$) and/or twisted in their orientations ($\Phi_n^{\mathrm{*F}}\neq\Phi_n^{\mathrm{*B}}$) between the forward-going and backward-going directions. Such FB-asymmetries are generic initial state long-range effects that should be carried over to the final state for $n=2$ and $n=3$, as the collective responses are expected to be linear for ellipticity and triangularity, $\vec{v}_2(\eta)\propto \vec{\epsilon}^{\mathrm{tot}}_2(\eta)$ and $\vec{v}_3(\eta)\propto \vec{\epsilon}^{\mathrm{tot}}_3(\eta)$.

To find out weather these initial state effects can survive the collective expansion, the AMPT model is used. This model has both the fluctuating initial geometry and final state flow generated by parton transport. Events are divided into several classes with different combinations of the asymmetry in $\epsilon_n^{\rm F}-\epsilon_n^{\rm B}$ and twist angle in $\Phi_n^{\mathrm{*F}}-\Phi_n^{\mathrm{*B}}$, and the flow harmonics are then calculated as a function of $\eta$ relative to the participant planes or raw event planes for each event class. The FB asymmetry and twist of the eccentricity vectors in $\eta$ are found to turn into similar FB asymmetry and twist in the final state flow vector $\vec{v}_n(\eta)$. The extracted asymmetry and twist are found to be nearly independent of the reference plane angle used, i.e. they are found to be the same among the three participant planes ($\Phi_n^{\mathrm{*F}},\Phi_n^{\mathrm{*B}}$ and $\Phi_n^{\mathrm{*}}$) and three event planes defined in different rapidity ranges. These observations strongly imply that initial longitudinal fluctuations are converted into similar longitudinal fluctuations in the final state harmonic flow. This is also supported by a recent full 3+1D ideal hydrodynamic calculation that use the AMPT initial condition in Ref.~\cite{Pang:2012uw,Roy}. The calculation reveals strong event-plane decorrelation very similar in detail to what was observed in full AMPT simulation. 

If similar effects are observed in the data analysis, it has the potential to greatly improve our understanding of the space-time evolution of the heavy ion collisions. The system would not be boost invariant event-by-event for harmonic flow, even though the flow rapidity distribution may appear boost-invariant when averaged over many events. Similar violation of boost-invariance for particle multiplicity has been suggested by previous experimental analysis and theoretical investigations~\cite{Bialas:2011bz,Bzdak:2012tp}, but the boost-invariance effect for flow could be much bigger. In fact if the $\eta$ dependence of the flow fluctuations can be measured, it could be used to constrain mechanisms for $\eta$-dependence of particle production (for example the $\alpha(\eta)$ and emission function $f^{\mathrm{F/B}}(\eta)$ via Eq.~\ref{eq:5}).

If the event-plane decorrelations arising from these initial state effects are large, we expect a breaking of the factorization of the flow harmonics for two-particle correlation into the flow harmonics obtained from the single particle distributions. Based on Figs.~\ref{fig:ecc1} (c)-(d), the decorrelation effects are expected to be large for $v_2$ in central collisions, and larger for $v_3$ across the full centrality range, where $\vec{\epsilon}_n^{\mathrm{F/B}}$ are dominated by random fluctuations. The decorrelations are also expected to be much stronger in p+A collisions and asymmetric collisions such as Cu+Au, and with very different rapidity dependence from A+A collisions. Future experimental searches for such long-range EbyE effects at RHIC and the LHC, using the event-shape selection~\cite{Schenke:2012fw,Huo:2013qma} and event-shape twist~\cite{Jia:2014vja} method, may provide new insights on the nature of initial state fluctuations and their relation to the dynamics of particle production in heavy ion collision.

We appreciate fruitful discussions with R.~Lacey, M.L.~Zhou, S.~Krishnan and N.N.~Ajitanand. This research is supported by NSF under grant number PHY-1305037.\bibliography{twistv3}{}
\bibliographystyle{apsrev4-1}
\appendix
\section{Shift of the center-of-mass and its influence on ellipticity}
\label{sec:app1}

\begin{figure}[t]
\begin{center}
\includegraphics[width=0.8\columnwidth]{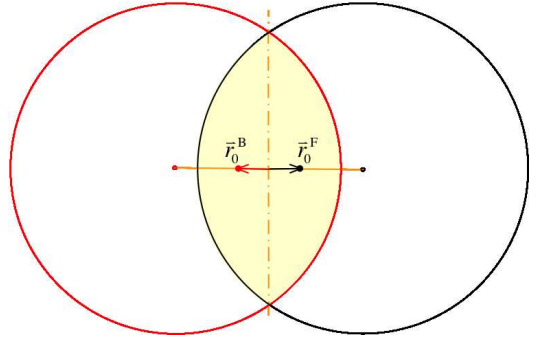}
\end{center}
\caption{\label{fig:app1} Schematic illustration of the origin for the shift of the center-of-mass between the two nuclei in the overlap region. Due to the almond shape, $\vec{\epsilon}^{\mathrm {F/B}}_2$ strongly aligns with the $\vec{r}_0^{\mathrm{F/B}}$.}
\end{figure}

In this section, we show that the difference between $\epsilon_2^{\mathrm{F/B}}$ and $\epsilon_2$ seen in Fig.~\ref{fig:ecc1}(a) is mainly due to an offset between the center-of-masses of the wounded nucleons in the two nuclei. As illustrated in Fig.~\ref{fig:app1}, the density distribution of the wounded nucleons in each nucleus is not uniform in the overlap region. This leads to a shift of the center-of-mass towards the center of the each nucleus. The shift vector $\vec{r}_0 = \langle re^{i\phi}\rangle =r_0e^{i\phi_0}$ for the two nuclei satisfy the following constraint:
\begin{eqnarray}
\npartf\vec{r}_0^{\rm F}+\npartb\vec{r}_0^{\rm B}=0
\end{eqnarray}

Since the energy profile of the fireball at pseudorapidity $\eta$ receives contributions from the two nuclei (Eq.~\ref{eq:4}), the shift vector for the fireball should interpolate between $\vec{r}_0^{\rm B}$ in the far backward rapidity to $\vec{r}_0^{\rm F}$ in the very forward rapidity,
\begin{eqnarray}
\nonumber
\vec{r}_0(\eta) &\equiv& r_0(\eta)e^{i\phi_0}=\alpha(\eta)\vec{r}_0^{\rm F}+(1-\alpha(\eta))\vec{r}_0^{\rm B}\\\label{eq:app11}
&=& (\alpha(\eta)(1+c)-c)\vec{r}_0^{\rm F}\;,
\end{eqnarray}
where $c = \npartf/\npartb$ is a constant.

The elliptic eccentricity vector can be expressed as:
\begin{eqnarray}
\vec{\epsilon}_2 = \left(\frac{\langle y^2-x^2\rangle}{\langle y^2+x^2\rangle},i\frac{4\langle xy\rangle}{\langle y^2+x^2\rangle}\right) 
\end{eqnarray}
A small global shift of $\vec{r}\rightarrow \vec{r}+\vec{r}_0(\eta)$ leads to a small correction of eccentricity vector:
\begin{eqnarray}
\label{eq:app12}
\delta\vec{\epsilon}_2 \approx -\frac{r_0^2(\eta)}{\langle r^2\rangle}\left(e^{2i\phi_0}+\vec{\epsilon}_2\right)
\end{eqnarray}
We can always chose the reference frame such that $\phi_0=0$ (by defining the $x$-axis along the impact parameter), and the correction to the eccentricities of the two nuclei at $\eta$ is obtained from Eqs.~\ref{eq:app11} and \ref{eq:app12}:
\begin{eqnarray}
\label{eq:app13}
\delta\vec{\epsilon}_2^{\rm F} &\approx& -(1+c)^2(1-\alpha)^2\left(1+\vec{\epsilon}_2^{\rm F}\right)\frac{\left(r_0^{\rm F}\right)^2}{\langle r^2\rangle^{\rm F}}\\
\delta\vec{\epsilon}_2^{\rm B} &\approx& -(1+c)^2\alpha^2\left(1+\vec{\epsilon}_2^{\rm B}\right)\frac{\left(r_0^{\rm F}\right)^2}{\langle r^2\rangle^{\rm B}}
\end{eqnarray}
From this we obtain the total correction to the eccentricity vector at $\eta$:
\begin{eqnarray}
\nonumber
\delta\vec{\epsilon}_2^{\mathrm{tot}}(\eta) &=& \alpha \delta\vec{\epsilon}_2^{\rm F} + (1-\alpha)\delta\vec{\epsilon}_2^{\rm B}\\\nonumber
&\approx& -4\alpha(1-\alpha)\left[1+(1-\alpha)\vec{\epsilon}_2^{\rm F}+\alpha\vec{\epsilon}_2^{\rm B}\right]\frac{\left(r_0^{\rm F}\right)^2+\left(r_0^{\rm B}\right)^2}{\langle r^2\rangle^{\rm F}+\langle r^2\rangle^{\rm B}}\\\label{eq:app14}
&\approx&-4\alpha(1-\alpha)\frac{\left(r_0^{\rm F}\right)^2+\left(r_0^{\rm B}\right)^2}{\langle r^2\rangle^{\rm F}+\langle r^2\rangle^{\rm B}}
\end{eqnarray}
where we have assumed $c=1$ and $r_0^{\rm F}\approx r_0^{\rm B}$ and $\langle r^2\rangle^{\rm F}\approx\langle r^2\rangle^{\rm B}$. This correction vanishes in forward and backward region where $\alpha=1$ or 0 as expected, and its value at mid-rapidity where $\alpha=1/2$ accounts for the differences between $\epsilon_2^{\mathrm{F/B}}$ and $\epsilon_2$, i.e.:
\begin{eqnarray}
\label{eq:app15}
\epsilon_2 \approx \epsilon_2^{\mathrm{F/B}} -\frac{\left(r_0^{\rm F}\right)^2+\left(r_0^{\rm B}\right)^2}{\langle r^2\rangle^{\rm F}+\langle r^2\rangle^{\rm B}}
\end{eqnarray}

\begin{figure}[t]
\begin{center}
\includegraphics[width=1\columnwidth]{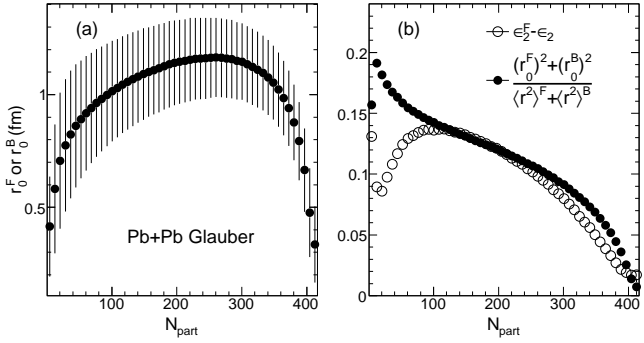}
\end{center}
\caption{\label{fig:app2} (a) The magnitude of the shift of the center-of-mass of wound nucleons in one nucleus from center-of-mass of all wound nucleons in the collisions. The bars represent the RMS values of the shift for events with the same $\npart$. (b) The $\epsilon_2^{\mathrm{F}}- \epsilon_2$ is compared with the estimate obtained form the shift (Eq.~\ref{eq:app14}). }
\end{figure}

We verified this relation explicitly using a monte-carlo Glauber model simulation and the results for Pb+Pb collisions at LHC energy are shown in Fig.~\ref{fig:app2}. The amount of shift, $r_0^{\mathrm{F}}$ or $r_0^{\mathrm{B}}$, is centrality dependent and reaches maximum of about 1.1~fm for mid-central collisions. The calculated correction on the eccentricity indeed largely accounts for the difference between $\epsilon_2^{\mathrm{F/B}}$ and $\epsilon_2$ (Fig.~\ref{fig:app2}(b)).

\section{Event plane resolution correction in light of the forward-backward twist}
\label{sec:app2}
The event-plane resolution correction is usually obtained via two-subevents method or three-subevents method~\cite{Poskanzer:1998yz} involving subevents A,B and C that do not use the same particle (this discussion is also applicable for the scalar-product method~\cite{Luzum:2012da}):
\begin{eqnarray}
\label{eq:app1}
 {\mathrm{Res}}^{\mathrm{2SE}}\{k\Psi^{\mathrm A}_{n}\}&=&\sqrt{\left\langle {\cos\Delta\Psi^{AB}_n} \right\rangle}\\\label{eq:app1b}
 {\mathrm{Res}}^{\mathrm{3SE}}\{k\Psi^{\mathrm A}_{n}\}&=&\sqrt{\frac{\left\langle {\cos\Delta\Psi^{AB}_n} \right\rangle\left\langle {\cos \Delta\Psi^{AC}_n}\right\rangle}{\left\langle {\cos \Delta\Psi^{BC}_n} \right\rangle}}.\\\nonumber
\end{eqnarray}
where $k$ is integer multiple of $n$, and $\Delta\Psi^{AB}_n = k \left(\Psi_n^{\mathrm A} - \Psi_n^{\mathrm B}\right)$ etc. The three subevents are usually selected from different pseudorapidity ranges. In the two subevents methods, the subevent A and B should have identical resolution typically achieved by choosing them from symmetric $\eta$ regions. These formulas assume that the true event plane angle in the three subevents are the same i.e. $\Phi_n^{\mathrm A}=\Phi_n^{\mathrm B}=\Phi_n^{\mathrm C}$, but clearly it is not the case when the EP angle is twisted continuously along the pseudorapidity. 

Assuming the event planes for subevent A and B are rotated relative to each other by
$\delta^{\mathrm{AB}}$ i.e. $\Delta\Psi^{AB}_n\rightarrow\Delta\Psi^{AB}_n+k\delta^{\mathrm{AB}}$. Then we have:
\begin{eqnarray}
\label{eq:app2}
\langle\cos(\Delta\Psi^{AB}_n+k\delta^{\mathrm{AB}})\rangle = \left\langle {\cos\Delta\Psi^{AB}_n} \right\rangle\cos k\delta^{\mathrm{AB}}
\end{eqnarray}
The second term in the expansion drops out, since $\left\langle {\sin\Delta\Psi^{AB}_n} \right\rangle=0$ when averaged over many events. The rotation angle is a smoothly varying function of $\eta$ and $\delta^{\mathrm{AC}}=\delta^{\mathrm{AB}}+\delta^{\mathrm{BC}}$. In this case, the resolution factor is modified as 
\begin{eqnarray}
\label{eq:app3}
{\mathrm{Res}}\{k\Psi^{\mathrm A}_{n}\}\rightarrow {\mathrm{Res}}\{k\Psi^{\mathrm A}_{n}\}\sqrt{r_k},
\end{eqnarray}
where correction factor $r_k$ are:
\begin{eqnarray}
\nonumber
r_k^{\mathrm{2SE}} = \cos\left[k\delta^{\mathrm{AB}}\right], r_k^{\mathrm{3SE}} = \frac{\cos\left[k\delta^{\mathrm{AB}}\right]\cos\left[k\delta^{\mathrm{AC}}\right]}{\cos\left[k\delta^{\mathrm{BC}}\right]}\\\label{eq:app4}
\end{eqnarray}
for the two-subevents method and the three-subevents method, respectively.

If the direction of the twists are the same for all events, the sine and cosine component of the flow coefficients can be calculated from Eq.~\ref{eq:11}, and we can then obtain the $v_n$ as $v_n=\sqrt{(\vnc)^2+(\vns)^2}$. However, if the event class is defined as all events in a given centrality range, then the probabilities for positive and negative twist are the same and the $\vns$ cancels out. In order to include the contribution from $\vns$, we should also correct for the twist effect in the raw flow coefficients:
\begin{eqnarray}
\nonumber
v_n(\eta)= \frac{v_n^{\mathrm{c,raw}}(\eta)}{\mathrm{Res}\{ n\Psi_n^{\rm A} \}}\rightarrow v_n(\eta)=\frac{v_n^{\mathrm{c,raw}}(\eta)}{\mathrm{Res}\{ n\Psi_n^{\rm A} \}} \frac{\cos \left[n\delta(\eta)\right]}{\sqrt{r_n}}.\\\label{eq:app5}
\end{eqnarray}
where $\cos \left[n\delta(\eta)\right]$ account for the rotation of the true event plane between the $\eta$ of the particles and the $\eta$ of the subevent A. All the rotation angle here should be treated as their root-mean-square values for the event class.

Experiments have used many different combinations of the $\eta$ ranges for the particles and subevents. Here we discuss a few common cases. Many analysis determine the resolution using two-subevents methods where subevents A and B are symmetric in $\eta$ (for example the forward counters~\cite{Adare:2010sp,star:2013wf,ALICE:2011ab} or the forward calorimeters~\cite{Aad:2012bu,Chatrchyan:2013kba}). The $v_n^{\mathrm{c,raw}}(\eta)$ is often calculated at mid-rapidity relative to either the combined event-plane (A+B) or the event-plane from one-side (A or B). In the first case, $\delta(\eta)\sim0$ and the correction is simply $1/\sqrt{\cos\left[n\delta^{\mathrm{AB}}\right]}\approx 1+\frac{1}{4}\left(\delta^{\mathrm{AB}}\right)^2$; in the second case,  $\delta(\eta)\sim 0.5\delta^{\mathrm{AB}}$ and the required correction is smaller, $\cos \left[n\delta(\eta)\right]/\sqrt{\cos\left[n\delta^{\mathrm{AB}}\right]}\approx 1+\frac{1}{8}\left(\delta^{\mathrm{AB}}\right)^2$. This correction is expected to be small for $v_2$, but can be significant for $v_3$ for which the decorrelation effects are large.

The situation for three-subevent methods are more complicated. But in general, subevents A and B should be chosen to be close to each other in $\eta$, such that $\delta^{\mathrm{AB}}$ is small and $\delta^{\mathrm{AC}}\approx\delta^{\mathrm{BC}}$. In this case, $r_n\approx \cos\left[n\delta^{\mathrm{AB}}\right]\approx1$. By choosing the subevent B to be in between the subevent A and the particles used for the differential flow measurement, one could reduce or even eliminate the correlation in Eq.~\ref{eq:app5} all together. We emphasize that as long as the far subevent C still have genuine long-range correlations with subevents A and B, the three-subevents method can still work. This is true even if the true event plane in subevent C is opposite in phase from those for subevents A and B. 

In p+Pb and d+Au collisions where the decorrelation effects are expected to be very large, the three-subevents method has to be used. The subevents A and B should be selected on the Pb-going or Au-going side, such that they both are strongly aligned with $\vec{\epsilon}_n$ of the Pb or Au. The subevent C can be chosen near mid-rapidity or on the proton-going side. As long as subevent C is not in the proton fragmentation region, it should still be partially correlated with eccentricity of the nucleus, and the resolution correction for subevent A can be calculated reliably.
\end{document}